\documentclass[twocolumn,aps,preprintnumbers,amsmath,amssymb,sort&compress,nofootinbib
]{revtex4-1}

\usepackage{amsfonts,amsmath,amssymb,ascmac,bm,tensor}
\usepackage{comment}
\usepackage{ifpdf}
\usepackage{slashed}
\usepackage{color}
\usepackage{mathptmx}
\usepackage[utf8]{inputenc}

\usepackage{cancel}

\ifpdf
  \usepackage{graphicx}     
  \usepackage[bookmarksopen,colorlinks=true,linkcolor=dark_blue,citecolor=dark_red,urlcolor=dark_red]{hyperref}
\else     
  \usepackage[dvipdfmx]{graphicx}     
  \usepackage[dvipdfmx,bookmarksopen,colorlinks=true,linkcolor=dark_blue,citecolor=dark_red,urlcolor=dark_red]{hyperref}
\fi

\definecolor{red}{rgb}{1,0,0}
\definecolor{dark_red}{rgb}{0.6,0,0}
\definecolor{dark_blue}{rgb}{0,0,0.6}
\definecolor{dark_yellow}{rgb}{0.6,0.6,0}

\begin{document}


\title{
Higgs Scalaron Mixed Inflation
}

\author{Yohei Ema}
\affiliation{Department of Physics, Faculty of Science, The University of Tokyo}

\begin{abstract}
\noindent
We discuss the inflationary dynamics of a system with a non-minimal coupling between the Higgs and the Ricci scalar
as well as a Ricci scalar squared term. There are two scalar modes in this system, 
\textit{i.e.} the Higgs and the spin-zero mode of the graviton, or the scalaron.
We study the two-field dynamics of the Higgs and the scalaron during inflation, 
and clarify the condition where inflation is dominated by the Higgs/scalaron.
We also find that the cut-off scale at around the vacuum 
is as large as the Planck scale,
and hence there is no unitarity issue,
although there is a constraint on the couplings from the perturbativity of the theory
at around the vacuum.
\end{abstract}

\date{\today}
\maketitle
\preprint{UT 17-04}


\section{Introduction}
\label{sec:intro}
\setcounter{equation}{0}
After the observation of the cosmic microwave background (CMB) anisotropy, 
inflation plays a central role in the modern cosmology. 
It is usually assumed that inflation is caused by potential energy of a scalar field, or the inflaton,
but there is no candidate within the standard model (SM). 
Hence we need to go beyond the SM to cause inflation.
Among a variety of such inflation models, 
the Higgs-inflation~\cite{Futamase:1987ua,CervantesCota:1995tz,Bezrukov:2007ep} 
and the $R^2$-inflation~\cite{Starobinsky:1980te,Barrow:1983rx,Whitt:1984pd,Barrow:1988xh} models are intriguing
because of their minimality as well as consistency with the CMB observation.
In the Higgs-inflation model, the SM Higgs boson plays the role of the inflaton
thanks to a large non-minimal coupling to the Ricci scalar.
In the $R^2$-inflation model, a spin-zero component of the metric (or the scaralon)
obtains a kinetic term and plays the role of the inflaton once we introduce a Ricci scalar squared term
in the action. It is known that both models predict a similar value of the spectral index
which is in good agreement with the Planck observation~\cite{Ade:2015lrj}.
It is also attractive that both models predict the tensor-to-scalar ratio that may be detectable
in the future CMB experiment (CMB-S4)~\cite{Abazajian:2016yjj}.

In the actual analysis of these models, 
it is sometimes assumed that the Higgs or the scalaron is 
the only scalar degree of freedom during inflation.
In reality, however, the Higgs must always be there 
even if we consider the $R^2$-inflation model.
In addition, if we consider the Higgs-inflation model, 
the large non-minimal coupling of the Higgs to the Ricci scalar 
may radiatively induce a large Ricci scalar squared term~\cite{Salvio:2015kka,Calmet:2016fsr}
that makes the scalaron dynamical as well. Hence it is more realistic 
to consider the dynamics of both the Higgs and the scalaron simultaneously.\footnote{
	A similar study with an additional scalar field instead of the scalaron
	with a non-minimal coupling to the Ricci scalar has been 
	performed in literature.
	See, \textit{e.g.} Refs.~\cite{Salopek:1988qh,Lebedev:2011aq,GarciaBellido:2011de,Greenwood:2012aj,
	Kaiser:2013sna,White:2013ufa,Kallosh:2013daa,
	Schutz:2013fua,Ferreira:2016vsc,Karananas:2016kyt} 
	and references therein.
} In this paper we will thus study the Higgs-scalaron two-field inflationary dynamics,\footnote{
	An analysis in this direction is also performed in Ref.~\cite{Kannike:2015apa,Salvio:2015kka}
	although some aspects we discuss in this paper such as
	the unitarity/perturbativity and the implication of 
	the electroweak vacuum metastability are not addressed there.
	See also Refs.~\cite{Torabian:2014nva,Myrzakulov:2015qaa,Bamba:2015uxa,Kaneda:2015jma,
	Rinaldi:2015uvu,Myrzakulov:2016tsz} 
	as other treatments.
} and derive the parameter dependence of the inflationary predictions in our system.
We will clarify the quantitative condition where inflation is dominated by the Higgs or the scalaron.
In addition, we will address the unitarity structure of our system, 
which is much different from that of the Higgs-inflation.

The organization of this paper is as follows.
In Sec.~\ref{sec:multi}, we discuss the inflationary dynamics of the Higgs-scalaron two-field system.
We first study the dynamics analytically, 
and later confirm it by numerical calculation.
In Sec.~\ref{sec:cut-off}, we study the unitarity structure of this system.
We find that the cut-off scale of our system is as large as the Planck scale,
which is similar to the case of the $R^2$-inflation rather than the Higgs-inflation.
In Sec.~\ref{sec:metastable}, we concentrate on the dynamics of the Higgs
when the electroweak (EW) vacuum is metastable.
The last section~\ref{sec:sum} is devoted to the summary and discussions.

\section{Inflationary dynamics}
\label{sec:multi}
In this section, we study the two-field dynamics of the Higgs and the scalaron during inflation.

\subsection{Action in Jordan/Einstein frame}
We start from the following action in the Jordan frame:
\begin{align}
	S = \int d^4x \sqrt{-g_J}
	&\left[
	\frac{M_P^2}{2}\left(1+\frac{\xi_h h^2}{M_P^2}\right)R_J \right. \nonumber \\
	&\left.
	+ \frac{\xi_s}{4}R_J^2
	-\frac{1}{2}g^{\mu\nu}_J\partial_\mu h\partial_\nu h - \frac{\lambda_h}{4}h^4
	\right],
	\label{eq:action_J1}
\end{align}
where $g_{J\mu\nu}$ is the metric (with the ``almost-plus'' convention), $g_J$ is the determinant of the metric,
$R_J$ is the Ricci scalar, $M_P$ is the reduced Planck mass
and $h$ is the Higgs in the unitary gauge.
We add the subscript $J$ for the quantities in the Jordan frame.
We consider only the case
\begin{align}
	\xi_s,~ \lvert \xi_h \rvert \gg 1.
\end{align}
In particular, we concentrate on the case $\xi_s > 0$ since otherwise there is a tachyonic mode.
On the other hand, we do not specify the signs of $\xi_h$ and $\lambda_h$.
Concerning the sign of $\lambda_h$, 
the current measurement of the top and Higgs masses indicates that 
it becomes negative at a high energy region, resulting in
the metastable EW vacuum~\cite{Sher:1988mj,Arnold:1989cb,Anderson:1990aa,Arnold:1991cv,
Espinosa:1995se,Isidori:2001bm,Espinosa:2007qp,Ellis:2009tp,Bezrukov:2009db,EliasMiro:2011aa,Bezrukov:2012sa,
Degrassi:2012ry,Buttazzo:2013uya,Bednyakov:2015sca},
although the stable EW vacuum is also still allowed.
In view of this, we consider both $\lambda_h > 0$ and $\lambda_h < 0$ in this paper.

By introducing an auxiliary field $s$, the action~\eqref{eq:action_J1} 
is rewritten as~\cite{Whitt:1984pd,Maeda:1988ab}\footnote{
This choice of the dual description is unique up to the shift and the rescaling of the auxiliary field $s$. 
For more details, see App.~\ref{app:eqv}.
}
\begin{align}
	S = \int d^4x \sqrt{-g_J}
	&\left[
	\frac{M_P^2}{2}\left(1+\frac{\xi_h h^2 + \xi_s s}{M_P^2}\right)R_J \right. \nonumber \\
	&\left.
	- \frac{\xi_s}{4}s^2
	-\frac{1}{2}g^{\mu\nu}_J\partial_\mu h\partial_\nu h - \frac{\lambda_h}{4}h^4
	\right].
	\label{eq:action_J2}
\end{align}
Note that the variation with respect to $s$ gives
\begin{align}
	s = R_J,
\end{align}
and we restore the original action~\eqref{eq:action_J1} after substituting it to Eq.~\eqref{eq:action_J2}.
The field $s$ corresponds to a spin-zero mode of the graviton that is dynamical due to 
the presence of the Ricci scalar squared term.
We call it a ``scalaron'' in this paper.

First we perform the Weyl transformation to
obtain the action in the Einstein frame. We define the metric in the Einstein frame as
\begin{align}
	g_{\mu\nu} = \Omega^2 g_{J\mu\nu},~~~
	\Omega^2 = 1+\frac{\xi_h h^2 + \xi_s s}{M_P^2}.
\end{align}
The Ricci scalar is transformed as
\begin{align}
	R_J = \Omega^2\left[R + 3\Box \ln \Omega^2 
	- \frac{3}{2}g^{\mu\nu}\partial_\mu \ln \Omega^2 \partial_\nu \ln \Omega^2\right],
\end{align}
where $R$ and $\Box$ are the Ricci scalar and the d'Alembert operator constructed from $g_{\mu\nu}$, respectively.
The action now reads
\begin{align}
	S = \int d^4x \sqrt{-g}
	\left[
	\frac{M_P^2}{2}R - \frac{3M_P^2}{4}g^{\mu\nu}\partial_\mu \ln\Omega^2 \partial_\nu \ln\Omega^2 
	\right. \nonumber \\  \left.
	-\frac{g^{\mu\nu}}{2\Omega^2}\partial_\mu h\partial_\nu h - U(h, s)
	\right],
\end{align}
where the potential in the Einstein frame is given by
\begin{align}
	U(h, s) \equiv \frac{\lambda_h h^4 + \xi_s s^2}{4\Omega^4}.
\end{align}
We define a new field $\phi$ as
\begin{align}
	\frac{\phi}{M_P} &\equiv \sqrt{\frac{3}{2}}\ln \Omega^2.
\end{align}
It corresponds to the inflaton degree of freedom in our system.
By eliminating $s$ in terms of $\phi$, we finally obtain
\begin{align}
	S = \int d^4x \sqrt{-g}
	&\left[
	\frac{M_P^2}{2}R - \frac{1}{2}g^{\mu\nu}\partial_\mu\phi\partial_\nu\phi
	\right. \nonumber \\ &\left.
	-\frac{1}{2}e^{-\chi}g^{\mu\nu}\partial_\mu h\partial_\nu h
	- U(\phi, h)
	\right],
	\label{eq:action_E}
\end{align}
where the potential now reads
\begin{align}
	U(\phi, h) = \frac{1}{4}e^{-2\chi}
	\left[
	\lambda_h h^4 
	+ \frac{M_P^4}{\xi_s}\left(e^{\chi}-1-\frac{\xi_h h^2}{M_P^2}\right)^2
	\right],
	\label{eq:pot_exact}
\end{align}
and we have defined
\begin{align}
	\chi \equiv \sqrt{\frac{2}{3}}\frac{\phi}{M_P}.
\end{align}
This is the master action in our system.
Note that so far we have not used any approximation.
In the following, we study the inflationary dynamics of this action in the Einstein frame.

\subsection{Two-field dynamics}
Now we study the inflationary dynamics of the action~\eqref{eq:action_E}.
An analysis for a similar system is performed in Ref.~\cite{Lebedev:2011aq},
and we follow that procedure here.
The action~\eqref{eq:action_E} contains the kinetic mixing term 
between $\phi$ and $h$, and hence we define the following field $\tau$ to solve the mixing:
\begin{align}
	\tau \equiv \frac{s}{h^2}.
	\label{eq:field_redef}
\end{align}
Note that $\tau = 0$ corresponds to the pure Higgs-inflation,
while $\tau = \infty$ corresponds to the pure $R^2$-inflation.
The kinetic terms now read
\begin{align}
	\mathcal{L}_\mathrm{kin} 
	=&
	-\frac{1}{2}\left(1+\frac{1}{6\left(\xi_h + \xi_s \tau\right)}\frac{e^{\chi}}{e^\chi - 1}\right)\left(\partial \phi\right)^2 
	\nonumber \\
	&- \frac{M_P^2}{8}\frac{\xi_s^2(1-e^{-\chi})}{\left(\xi_h+ \xi_s \tau\right)^3}\left(\partial \tau\right)^2
	+ \frac{M_P}{2\sqrt{6}}\frac{\xi_s}{\left(\xi_h + \xi_s \tau\right)^2}(\partial \phi) (\partial \tau).
\end{align}
Since we are interested in the inflationary dynamics, we concentrate on the case
\begin{align}
	\xi_h h^2 + \xi_s s \gg M_P^2,
	~~\mathrm{or}~~
	e^\chi \gg 1,
	\label{eq:large_field}
\end{align}
in this section. 
Then, the kinetic terms are approximated as
\begin{align}
	\mathcal{L}_\mathrm{kin} 
	=&
	-\frac{1}{2}\left(1+\frac{1}{6\left(\xi_h + \xi_s \tau\right)}\right)\left(\partial \phi\right)^2 
	\nonumber \\
	&- \frac{M_P^2}{8}\frac{\xi_s^2}{\left(\xi_h+ \xi_s \tau\right)^3}\left(\partial \tau\right)^2
	+ \frac{M_P}{2\sqrt{6}}\frac{\xi_s}{\left(\xi_h + \xi_s \tau\right)^2}(\partial \phi) (\partial \tau).
\end{align}
Note that $\tau$ satisfies
\begin{align}
	\xi_h + \xi_s \tau \gg \frac{M_P^2}{h^2} > 0,
	\label{eq:tau_range}
\end{align}
when the condition~\eqref{eq:large_field} is satisfied,
and hence the kinetic term of $\tau$ has a correct sign.
In the following we assume
\begin{align}
	\xi_h + \xi_s \tau \gg 1,
	\label{eq:tau_range}
\end{align}
which is true for $h \lesssim M_P$.\footnote{
	It might not be true for, \textit{e.g.} the critical case~\cite{Hamada:2014iga,Bezrukov:2014bra} 
	where $\xi_h \sim \mathcal{O}(10)$ since the Higgs field value is 
	of order $M_P$ during inflation in that case.
} Then after defining the canonically normalized field $\tau_c$ as
\begin{align}
	d\tau_c \equiv \frac{\xi_sM_P}{2\left(\xi_h + \xi_s\tau\right)^{3/2}}d\tau,
\end{align}
the kinetic mixing term between $\phi$ and $\tau_c$ is suppressed by $1/\sqrt{\xi_h + \xi_s \tau}$. 
Therefore, to the leading order in it, 
we can approximate the kinetic term as
\begin{align}
	\mathcal{L}_\mathrm{kin}
	=
	-\frac{1}{2}\left(\partial \phi\right)^2 
	- \frac{M_P^2}{8}\frac{\xi_s^2}{\left(\xi_h \tau^2 + \xi_s\right)^3}\left(\partial \tau\right)^2.
\end{align}
There is no kinetic mixing between $\phi$ and $\tau$ to the zero-th order in 
$1/\sqrt{\xi_h + \xi_s \tau}$.

Armed with these diagonalized fields, we now study the structure of the potential,
which is expressed as
\begin{align}
	U(\phi, \tau) &= \frac{M_P^4}{4}\frac{\lambda_h + \xi_s\tau^2}{\left(\xi_h + \xi_s\tau\right)^2}
	\left[1-\exp\left(-\sqrt{\frac{2}{3}}\frac{\phi}{M_P}\right)\right]^2.
	\label{eq:pot}
\end{align}
To the leading order in Eq.~\eqref{eq:large_field}, 
its derivative gives
\begin{align}
	\frac{\partial U}{\partial \tau_c}
	&= \frac{-\lambda_h + \xi_h \tau}{\left(\xi_h + \xi_s\tau\right)^{3/2}}M_P^3.
\end{align}
Note that we take the derivative with respect to $\tau_c$, not $\tau$.
Hence the extrema are
\begin{align}
	\tau = \infty,~ \tau_\mathrm{min},
\end{align}
where we have defined
\begin{align}
	\tau_\mathrm{min} \equiv 
	\frac{\lambda_h}{\xi_h}.
	\label{eq:tau_min}
\end{align}
In particular, $\tau = 0$ is not an extremum of the potential. 
It means that the pure Higgs-inflation is \textit{never} realized in our system,
although the pure $R^2$-inflation is possible.
Nevertheless, there is some parameter region where inflation is caused mostly by the Higgs as we will see below.

In order to be the inflationary trajectory,
these extrema must be minima of the potential.
The second derivative at each extremum is given by
\begin{align}
	\left.\frac{\partial^2 U}{\partial \tau_c^2}\right\rvert_{\tau = \infty} 
	&= -\frac{\xi_h}{\xi_s}M_P^2, 
	\label{eq:second_deriv1} \\
	\left.\frac{\partial^2 U}{\partial \tau_c^2}\right\rvert_{\tau = \tau_\mathrm{min}}
	&= \frac{2\xi_h}{\xi_s}M_P^2.
	\label{eq:second_deriv2}
\end{align}
Note again that we take the derivative with respect to $\tau_c$, not $\tau$.
Thus, $\tau = \infty$ is the minimum for $\xi_h\xi_s < 0$, 
while $\tau = \tau_\mathrm{min}$ is the minimum for $\xi_h\xi_s>0$.
Also, the potential at the minimum must be positive to cause inflation.
The potential at each extremum is given by
\begin{align}
	\left.\frac{U}{M_P^4}\right\rvert_{\tau = \infty} &= \frac{1}{4\xi_s}, \\
	\left.\frac{U}{M_P^4}\right\rvert_{\tau = \tau_\mathrm{min}} &= \frac{1}{4\left(\xi_h^2/\lambda_h + \xi_s\right)}.
	\label{eq:pot_min}
\end{align}
The former is always positive in the case of our interest,
while the latter together with Eq.~\eqref{eq:tau_range} gives a non-trivial constraint on the parameters.
In fact, Eq.~\eqref{eq:tau_range} in particular means
\begin{align}
	\xi_h + \xi_s \tau_\mathrm{min} > 0,
	\label{eq:taumin_range}
\end{align}
and hence we obtain the condition
\begin{align}
	\lambda_h \xi_h > 0,
\end{align}
by combining the requirement that Eq.~\eqref{eq:pot_min} is positive.
Thus, the trajectory with $\tau = \tau_\mathrm{min}$ causes inflation only if
$\xi_h > 0$ and $\lambda_h > 0$. Note that we consider only the case $\xi_s > 0$.\footnote{
	It is sufficient to require only these conditions since Eq.~\eqref{eq:taumin_range}
	is trivially satisfied under them.
}
 
In summary, there are three cases that are relevant for inflation: (a)~$\xi_h > 0,~ \xi_s > 0$ and $\lambda_h > 0$,
(b)~$\xi_h < 0,~ \xi_s>0$ and $\lambda_h>0$, and (c)~$\xi_h<0,~\xi_s>0$ and $\lambda_h<0$.
From now we mainly concentrate on the case~(a) since $\tau = \tau_\mathrm{min}$ is the minimum only in this case.
We briefly comment on the case~(b) in the end of this subsection. 
The case~(c) is also interesting since it corresponds to the metastable EW vacuum. 
Hence we discuss the case~(c) in detail in Sec.~\ref{sec:metastable}.

\subsubsection*{Case~(a): $\xi_h>0,~~\xi_s>0$~ and~ $\lambda_h>0$}

\begin{figure}
\begin{center}
\includegraphics[scale=0.4]{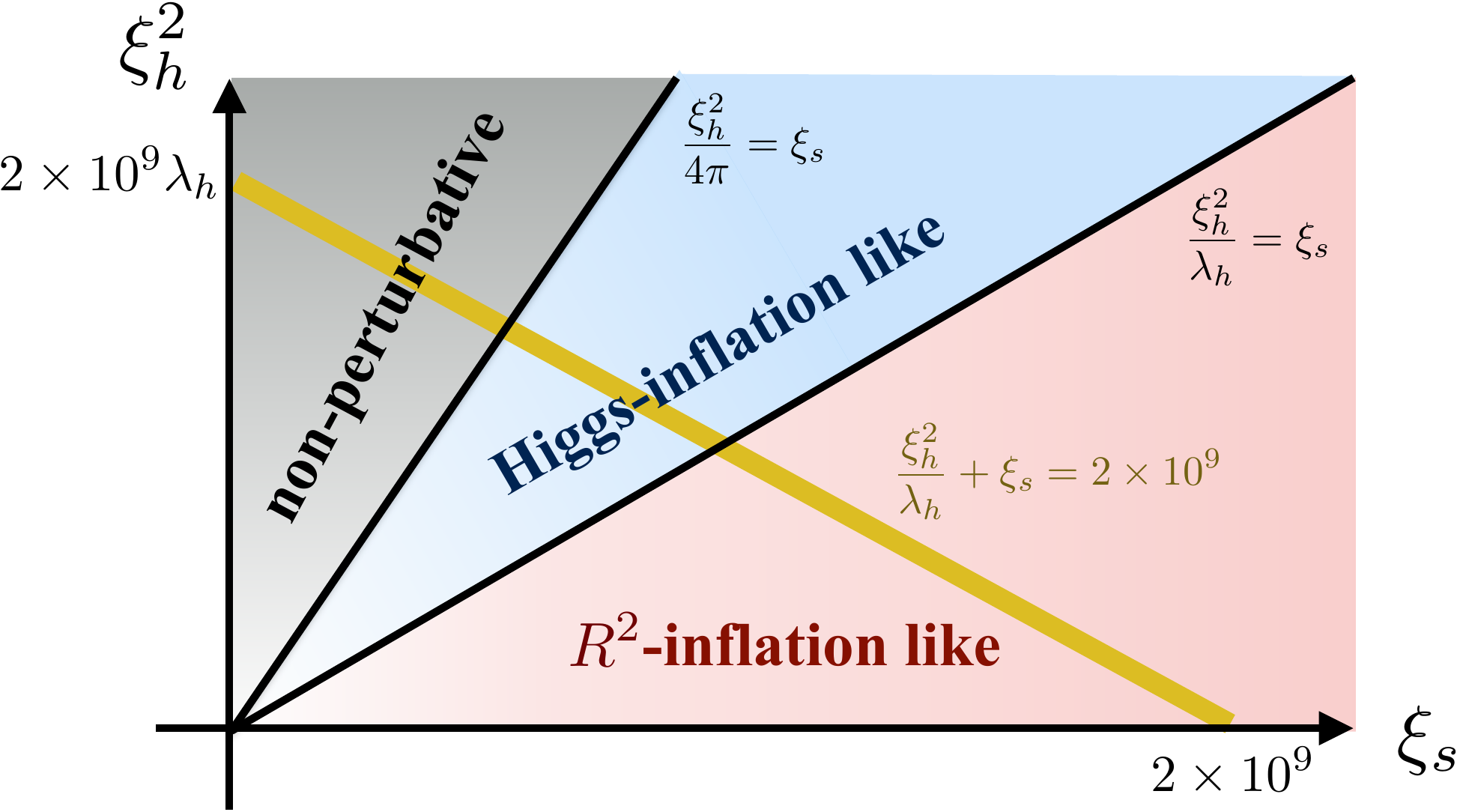}
\end{center}
\caption {
	Schematic picture of the parameter region in the case~(a).
	The inflaton is dominated by the scalaron for $\xi_h^2/\lambda_h \lesssim \xi_s$
	(the \textcolor{dark_red}{red region}), while it is dominated by the Higgs 
	for  $\xi_h^2/\lambda_h \gtrsim \xi_s$ (the \textcolor{dark_blue}{blue region}).
	The \textcolor{dark_yellow}{yellow band} corresponds to the parameter region 
	consistent with the CMB observation. 
	If $\xi_h^2/\xi_s \gtrsim 4\pi$ (the black region), 
	the UV theory is non-perturbative so that
	we might not obtain the standard model effective theory 
	even below the energy scale of the inflaton mass. 
	For more details on this point, see Sec.~\ref{sec:cut-off}.
}
\label{fig:schematic}
\end{figure}

In this case, the potential minimum for $\tau$ is given by
\begin{align}
	\tau = \tau_\mathrm{min} \neq 0,
\end{align}
and hence the inflaton $\phi$ is a mixture of the Higgs and the scalaron.
We first assume that $\tau$ sits at this minimum during inflation.
Later we will see that this assumption is actually valid.
Recalling that $\tau = s/h^2$, for $\xi_h \gg \xi_s \tau_\mathrm{min}$, the inflaton is dominated by the Higgs,
while in the other limit, it is mostly composed of the scalaron. Thus, the situation is as follows:\footnote{
	A similar condition is obtained for a scale invariant model with an additional dilaton field
	in Ref.~\cite{Gorbunov:2013dqa}. 
	It may be reasonable because the theory is almost scale invariant in our case
	as long as we consider the inflationary dynamics.
	It is also consistent with the rough estimation in Ref.~\cite{Barbon:2015fla}.
}
\begin{align}
	\begin{cases}
	\lambda_h \xi_s \ll \xi_h^2: & \mathrm{Higgs\text{-}inflation~like}, \vspace{2mm}\\
	\lambda_h \xi_s \gg \xi_h^2: & R^2\mathrm{\text{-}inflation~like}.
	\end{cases}
	\label{eq:the_cond}
\end{align}
It is also seen from the potential. The potential for $\tau = \tau_\mathrm{min}$ is given by
\begin{align}
	\left.U\right\rvert_{\tau = \tau_\mathrm{min}} 
	= \frac{M_P^4}{4}\frac{1}{\xi_h^2/\lambda_h + \xi_s}\left[1-\exp\left(-\sqrt{\frac{2}{3}}\frac{\phi}{M_P}\right)\right]^2,
	\label{eq:pot_inf}
\end{align}
and hence it is the same as the Higgs-inflation for $\lambda_h\xi_s \ll \xi_h^2$, 
while the same as the $R^2$-inflation for $\lambda_h\xi_s \gg \xi_h^2$.
In the intermediate case, the inflaton is a mixture of the Higgs and the scalaron,
which we call a ``Higgs scalaron mixed inflation.''
In order to reproduce the normalization of the CMB anisotropy, 
the parameters should satisfy~\cite{Ade:2015lrj}
\begin{align}
	\frac{\xi_h^2}{\lambda_h} + \xi_s \simeq 2\times 10^{9}.
	\label{eq:norm_CMB}
\end{align}
It is well-known that this type of model is in good agreement with the spectral index
observed by the CMB experiments.
In Fig.~\ref{fig:schematic}, we show the schematic picture of the parameter region in the case~(a).

Now we investigate the assumption that $\tau$ sits at the minimum of its potential during inflation.
It is valid if the mass squared of $\tau_c$ at around the minimum 
$\sim \xi_h M_P^2/\xi_s$ (see Eq.~\eqref{eq:second_deriv2})
is much larger than the Hubble parameter squared during inflation $H_\mathrm{inf}^2$.
The ratio is estimated as
\begin{align}
	\frac{\xi_h M_P^2/\xi_s}{H_\mathrm{inf}^2} 
	&\sim \frac{\xi_h}{\xi_s}\left(\frac{\xi_h^2}{\lambda_h} + \xi_s\right) > \xi_h \gg 1.
	\label{eq:tau_heavy}
\end{align}
Thus, $\tau$ sits at the minimum of its potential during inflation as long as $\xi_h \gg 1$ is satisfied,
which is the case of our interest, and hence we have verified our assumption.
It means that the inflationary dynamics effectively reduces to a single field case, 
whose potential is given by Eq.~\eqref{eq:pot_inf}.

\subsubsection*{Case~(b): $\xi_h<0,~~\xi_s>0$~ and~ $\lambda_h>0$}
In this case, the potential minimum for $\tau$ is
\begin{align}
	\tau = \infty,
\end{align}
and hence inflation is caused solely by the scalaron. 
Actually, the potential is given by
\begin{align}
	\left.U\right\rvert_{\tau = \infty} 
	= \frac{M_P^4}{4\xi_s}\left[1-\exp\left(-\sqrt{\frac{2}{3}}\frac{\phi}{M_P}\right)\right]^2,
\end{align}
which is nothing but the potential of the $R^2$-inflation model.
It is consistent with the CMB for $\xi_s \simeq 2\times 10^{9}$~\cite{Ade:2015lrj}.

\subsection{Numerical confirmation}
In this subsection, we perform numerical calculation to confirm the analysis in the previous subsection.
In particular, we numerically study the case~(a), or the Higgs scalaron mixed inflation case.
While some approximations are used in the previous subsection, 
we emphasize that we use \textit{no approximation} in this subsection.
More explicitly, we directly solve the background equations of motion 
derived from the action~\eqref{eq:action_E}, which read
\begin{align}
	0 &= \ddot{\phi} + 3H\dot{\phi} + \frac{e^{-\chi}}{\sqrt{6}M_P}\dot{h}^2 + \frac{\partial U}{\partial \phi}, \\
	0 &= \ddot{h} + \left(3H - \sqrt{\frac{2}{3}}\frac{\dot{\phi}}{M_P}\right)\dot{h} + e^\chi \frac{\partial U}{\partial h}, \\
	\dot{H} &= -\frac{1}{2M_P^2}\left(\dot{\phi}^2 + e^{-\chi}\dot{h}^2\right), \\
	H^2 &= \frac{1}{6M_P^2}\left(\dot{\phi}^2 + e^{-\chi}\dot{h}^2 + 2U\right),
	\label{eq:ene_cons}
\end{align}
where the potential is
\begin{align}
	U(\phi, h) = \frac{1}{4}e^{-2\chi}
	\left[
	\lambda_h h^4 
	+ \frac{M_P^4}{\xi_s}\left(e^{\chi}-1-\frac{\xi_h h^2}{M_P^2}\right)^2
	\right].
	\label{eq:pot_exact2}
\end{align}
Here we take the background metric in the Einstein frame to be
the Friedmann-Lema\^itre-Robertson-Walker (FLRW) one without spatial curvature, 
$H$ is the Hubble parameter and the dots denote the derivatives with respect to the time.
For the convenience of readers, we write down the explicit forms of the derivatives of the potential:
\begin{align}
	\frac{\partial U}{\partial \phi} 
	&= 
	\frac{e^{-2\chi}M_P^3}{\sqrt{6}\,\xi_s}
	\left[
	\left(1+\frac{\xi_h h^2}{M_P^2}\right)\left(e^\chi - 1 - \frac{\xi_h h^2}{M_P^2}\right)
	- \frac{\lambda_h \xi_s h^4}{M_P^4}
	\right], \\
	\frac{\partial U}{\partial h} 
	&=
	e^{-2\chi} \frac{M_P^2 h}{\xi_s}
	\left[
	-\xi_h\left(e^\chi -1 -\frac{\xi_h h^2}{M_P^2}\right)
	+ \frac{ \lambda_h \xi_s h^2}{M_P^2}
	\right].
\end{align}
Actually, these equations of motion are redundant, and hence we have used the last one~\eqref{eq:ene_cons}
to check the consistency of our numerical calculation. We have numerically solved the first three equations of motion, 
and checked that Eq.~\eqref{eq:ene_cons} is satisfied at least better than $10^{-5}$ level.

\begin{figure}[t]
\begin{center}
\includegraphics[scale=0.6]{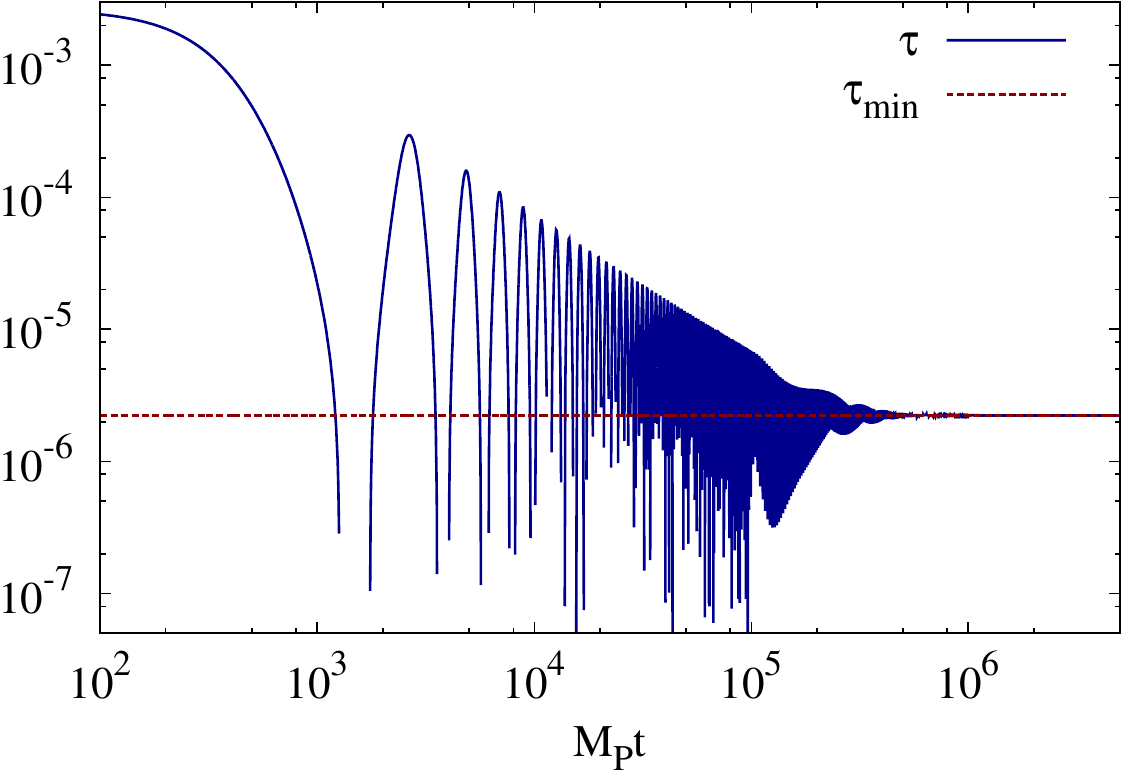}
\includegraphics[scale=0.6]{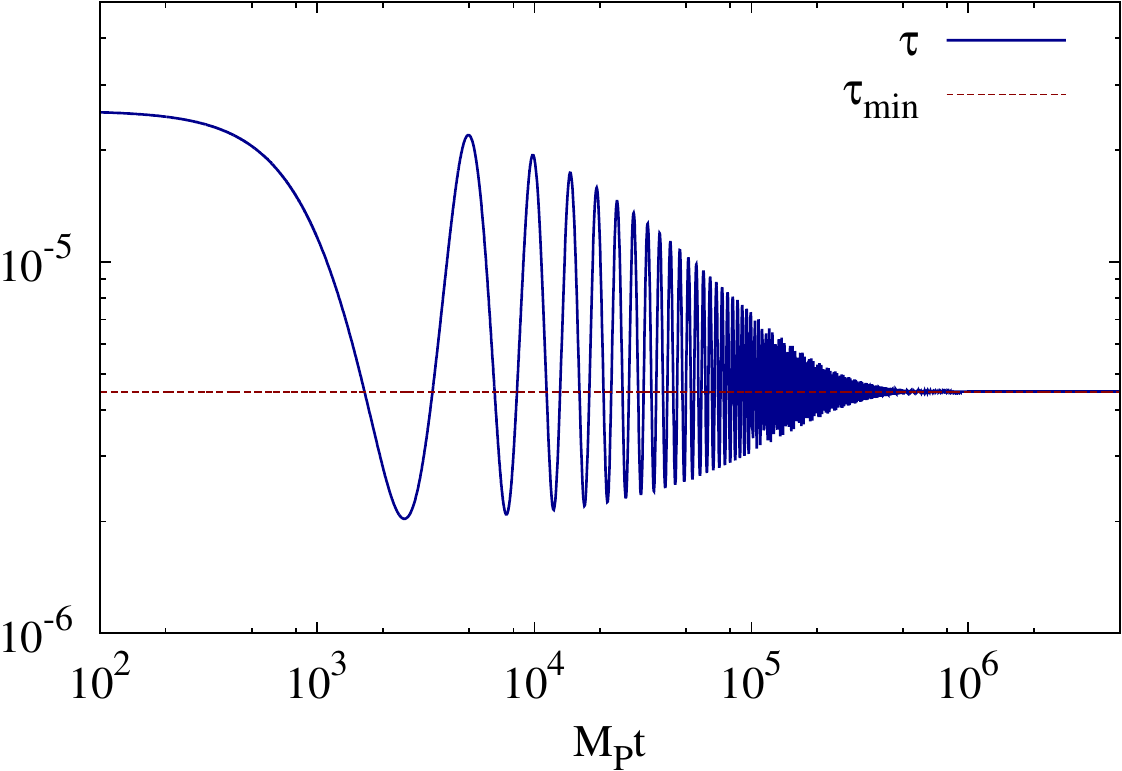}
\end{center}
\caption {
	The time evolution of $\tau$ calculated from $\phi$ and $h$.
	The x-axis is time in the unit of the Planck mass inverse, 
	and the y-axis is the values of $\tau$ and $\tau_\mathrm{min}$.
	The \textcolor{dark_blue}{blue line} corresponds to our numerical calculation, 
	while the \textcolor{dark_red}{red dashed line} is $\tau_\mathrm{min}$ given in Eq.~\eqref{eq:tau_min}.
	We take the parameters as follows.
	Top: $\xi_s = 5\times10^8$, $\xi_h^2 = 2\times10^7$, $\lambda_h = 0.01$, $\phi_\mathrm{ini} = 6\,M_P$,
	and $h_\mathrm{ini} = 0.01\,M_P$ (Higgs-inflation like). 
	Bottom: $\xi_s = 2\times10^{9}$, $\xi_h^2 = 5\times10^6$, $\lambda_h = 0.01$, $\phi_\mathrm{ini} = 6\,M_P$,
	and $h_\mathrm{ini} = 0.05\,M_P$ ($R^2$-inflation like). 
	Here $\phi_\mathrm{ini}$ and $h_\mathrm{ini}$ are the initial
	field values of the inflaton and Higgs, respectively. We take the initial velocities of the fields as zero.
	$\tau$ starts to oscillate when $\xi_hM_P^2 t^2/\xi_s \sim 1$. After hundreds or thousands of oscillations
	$\tau$ eventually settles down to its potential minimum $\tau_\mathrm{min}$. 
	Note that it happens within a few e-foldings
	because of the inequality~\eqref{eq:tau_heavy}, 
	and hence $\tau$ has almost no effect on the inflationary dynamics.
}
\label{fig:tau}
\end{figure}

\begin{figure}[t]
\begin{center}
\includegraphics[scale=0.65]{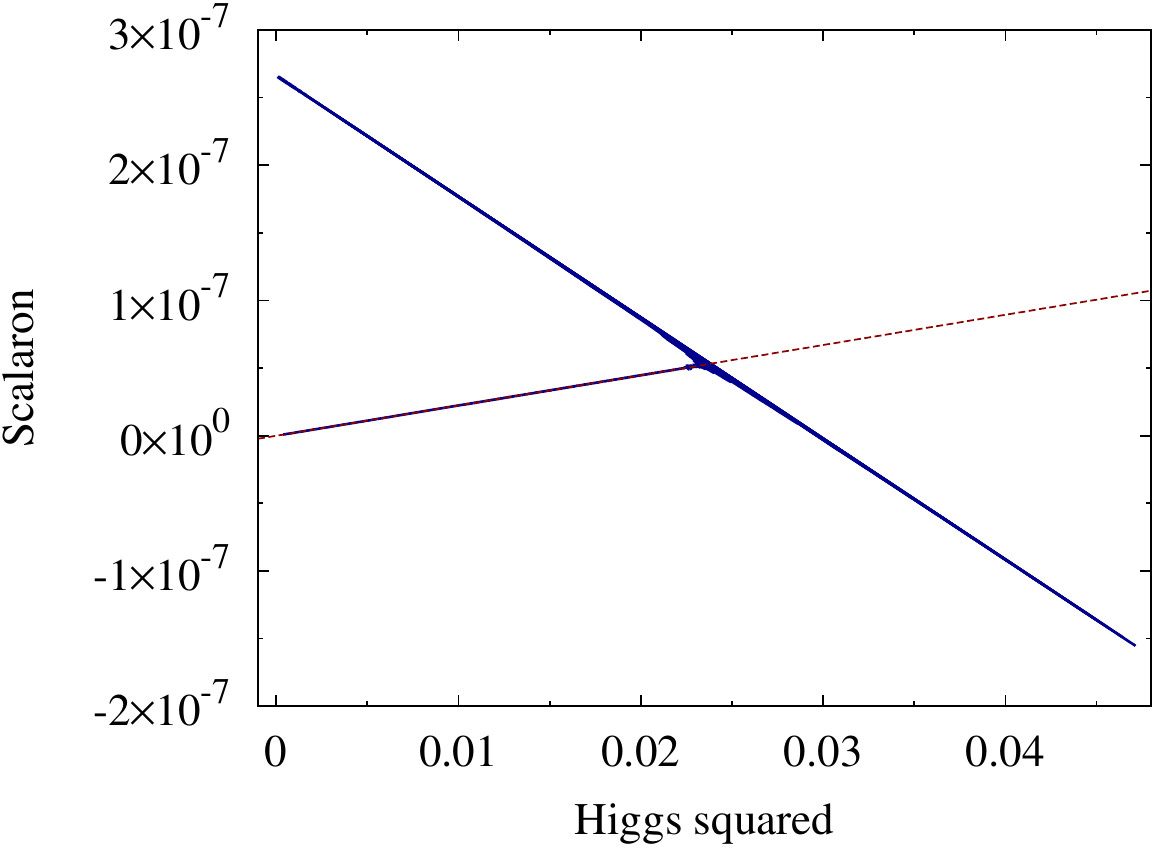}
\includegraphics[scale=0.65]{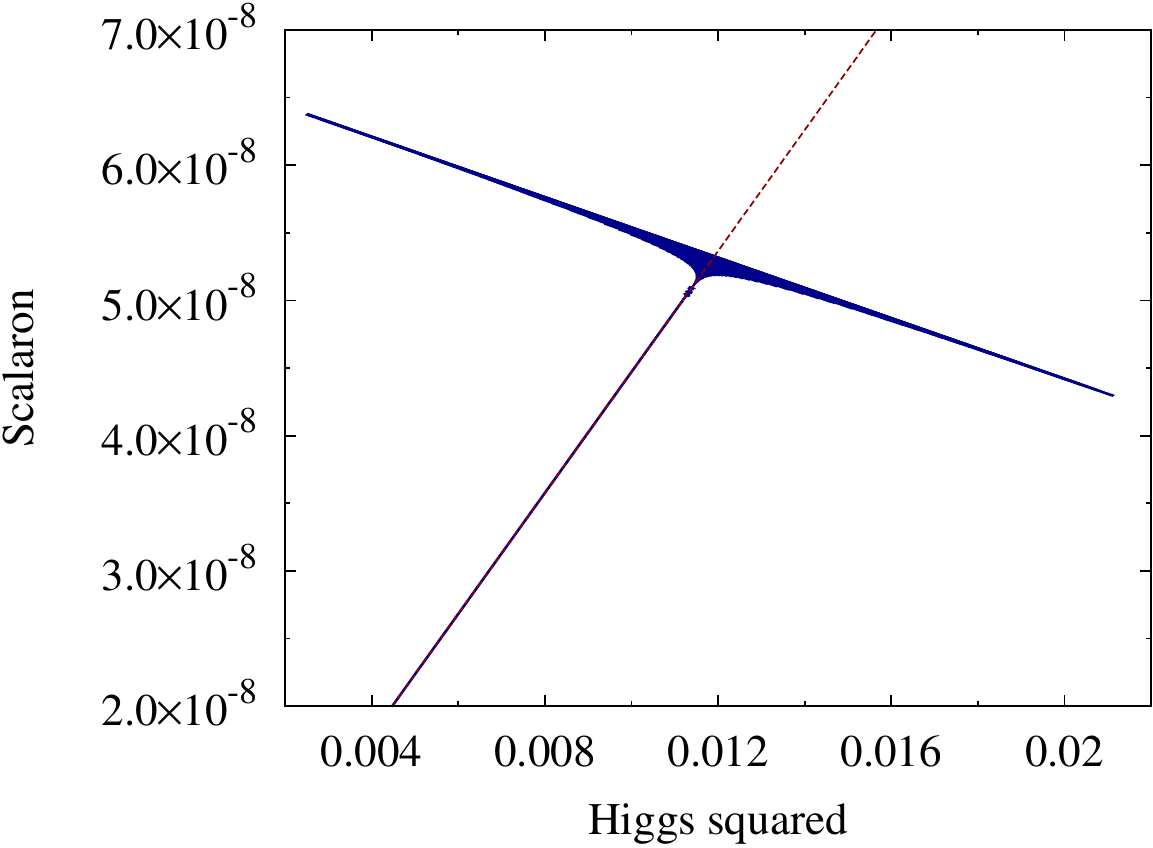}
\end{center}
\caption {
	The dynamics of the Higgs and scalaron in the field space.
	The x-axis is the Higgs squared $h^2$, 
	and the y-axis is the scalaron $s$ calculated from $\phi$ and $h$.
	We take $M_P = 1$ in this plot.
	The \textcolor{dark_blue}{blue line} is our numerical calculation, 
	while the \textcolor{dark_red}{red dashed line} is the direction of $\phi$ that is orthogonal to $\tau$.
	We take the parameters as follows.
	Top: $\xi_s = 5\times10^8$, $\xi_h^2 = 2\times 10^7$, $\lambda_h = 0.01$, $\phi_\mathrm{ini} = 6\,M_P$,
	and $h_\mathrm{ini} = 10^{-2}\,M_P$ (Higgs-inflation like).
	Bottom: $\xi_s = 2\times10^{9}$, $\xi_h^2 = 5\times10^6$, $\lambda_h = 0.01$, $\phi_\mathrm{ini} = 6\,M_P$,
	and $h_\mathrm{ini} = 0.05\,M_P$ ($R^2$-inflation like).
	Here $\phi_\mathrm{ini}$ and $h_\mathrm{ini}$ are the initial
	field values of the inflaton and Higgs, respectively. 
	We have taken the initial velocities of the fields as zero.
	We have followed the inflationary dynamics until $M_Pt = 2\times10^7$.
	The two fields oscillate very frequently in the $\tau$ direction at first,
	but soon $\tau$ settles down to its potential minimum $\tau_\mathrm{min}$, 
	and the two field dynamics follows
	the trajectory of $\phi$, which is expressed by the \textcolor{dark_red}{red line}.
	For the time evolution of $\tau$, see Fig.~\ref{fig:tau}.
}
\label{fig:phase_space}
\end{figure}

In Fig.~\ref{fig:tau}, we show the time evolution of $\tau$. 
The \textcolor{dark_blue}{blue line} corresponds to our numerical calculation, 
while the \textcolor{dark_red}{red dashed line} is $\tau_\mathrm{min}$ given in Eq.~\eqref{eq:tau_min}.
The top panel corresponds to the Higgs-inflation like case, 
while the bottom one does to the $R^2$-inflation like case. 
See the caption for more details on the parameters and the initial conditions.
As we can see from the figures, after hundreds or thousands of oscillations 
$\tau$ eventually settles down to its potential minimum. 
It happens within a few e-foldings, 
and hence $\tau$ has almost no effect on the inflationary dynamics at all.
This result confirms our analysis in the previous subsection.
In Fig.~\ref{fig:phase_space}, we also show the inflationary dynamics in the Higgs-scalaron field space.
The \textcolor{dark_blue}{blue line} again corresponds to our numerical calculation, 
while the \textcolor{dark_red}{red dashed line} is the $\phi$ direction that is orthogonal to the $\tau$ direction.
We can see that the fields oscillate in the $\tau$ direction at first, but after $\tau$ settles down to its potential minimum,
the system follows the trajectory of $\phi$. 
Thus, it is again consistent with our analytical treatment in the previous subsection.
In summary, we have verified that our analysis in the previous subsection well describes
the actual inflationary dynamics of the present system.

\section{Cut-off scale and perturbativity}
\label{sec:cut-off}
In this section, we discuss the cut-off scale and the perturbativity 
of our model at around the vacuum.
Note that the discussion in this section holds for all the cases~(a), (b) and~(c).
We start our discussion with the exact action in the Einstein frame~\eqref{eq:action_E}.
The minimum of the potential~\eqref{eq:pot_exact} is at $(\phi,\,h) = (0,\,0)$,
so we expand the action around that point:
\begin{align}
	S 
	&\simeq
	\int d^4x \sqrt{-g}
	\left[ \frac{M_P^2}{2}R - \frac{1}{2}\left(\partial \phi\right)^2 
	- \frac{1}{2}\left(1-\chi\right)\left(\partial h\right)^2
	\right. 
	\nonumber \\ 
	&\left.
	-\frac{\lambda_\mathrm{UV}}{4}\left(1-2\chi\right)h^4
	-\frac{m_\phi^2}{2}\phi^2 +\frac{m_\phi}{3\sqrt{2\xi_s}} \phi^3
	\right.
	\nonumber \\
	&\left.
	-\frac{7}{108\xi_s}\left(1-\frac{3\chi}{7}\right)\phi^4
	+\frac{\xi_h m_\phi}{\sqrt{2\xi_s}} \phi h^2
	-\frac{\xi_h}{2\xi_s}\left(1-\frac{7\chi}{9}\right)\phi^2h^2
	\right],
\end{align}
where we have defined the Higgs quartic coupling in the original ultraviolet (UV) theory
and the inflaton mass squared as\footnote{
	Although $\lambda_\mathrm{UV}$ is shifted due to the mixing between $\phi$ and $h$, 
	it does not help to stabilize the EW vacuum for $\lambda_h < 0$.
	This is because the CMB observation fixes $m_\phi \sim 10^{13}\,\mathrm{GeV}$,
	which is much higher than the instability scale of the Higgs potential~\cite{Lebedev:2012zw,EliasMiro:2012ay}.
	Note that inflation is dominated by the scalaron for $\lambda_h < 0$,
	and hence $m_\phi$ is fixed.
}
\begin{align}
	\lambda_\mathrm{UV} &\equiv \lambda_h + \frac{\xi_h^2}{\xi_s}, \\
	m_\phi^2 &\equiv \frac{M_P^2}{3\xi_s}.
\end{align}
Neglected terms are suppressed by higher powers of $M_P$.
Thus, recalling that $\chi = \sqrt{2/3}\;\phi/M_P$, we can see that 
the cut-off scale of our model is as high as the Planck scale.
It is in contrast to the case of the Higgs-inflation,
where a power counting argument\footnote{
	Analysis beyond the power counting might change the situation~\cite{Calmet:2013hia,Escriva:2016cwl}.
} suggests that the cut-off scale $\Lambda_{\mathrm{cut};h}$
at around the vacuum is of 
$\mathcal{O}(M_P/\xi_h)$~\cite{Burgess:2009ea,Barbon:2009ya,
Burgess:2010zq,Hertzberg:2010dc,Kehagias:2013mya}.\footnote{
	It does not necessarily mean an inconsistency of the Higgs-inflation during inflation~\cite{Bezrukov:2010jz}
	although the unitarity can be broken during the inflaton oscillation epoch~\cite{Ema:2016dny}.
	See, \textit{e.g.} Refs.~\cite{Giudice:2010ka,Lerner:2010mq,Barbon:2015fla} 
	for construction of a UV completed model and
	also
	Refs.~\cite{Burgess:2014lza,Fumagalli:2016lls} for the discussion of UV effects on the Higgs-inflation.
} Rather, our model is similar to the case of the $R^2$-inflation.
In the $R^2$-model, the cut-off scale is the Planck scale because the scalaron is just an auxiliary field 
in the Jordan frame, and hence it can absorb the large non-minimal coupling 
with the curvature~\cite{Hertzberg:2010dc,Kehagias:2013mya}.
Similarly in our case, the scalaron absorbs the large non-minimal couplings $\xi_s$ and $\xi_h$
so that it can avoid the unitarity issue.
An interesting point of our model is that, Higgs can be the dominant component of the inflaton while keeping
the cut-off to be the Planck scale.

Still, there is a constraint on the parameters if we require the theory to be perturbative.
In order to see this, let us keep only the renormalizable terms in the action at around the vacuum:
\begin{align}
	S 
	\simeq
	\int d^4x \sqrt{-g}
	&\left[ \frac{M_P^2}{2}R - \frac{1}{2}\left(\partial \phi\right)^2 
	- \frac{1}{2}\left(\partial h\right)^2
	\right. 
	\nonumber \\ 
	&\left.
	-\frac{\lambda_\mathrm{UV}}{4}h^4
	-\frac{m_\phi^2}{2}\phi^2 +\frac{m_\phi}{3\sqrt{2\xi_s}} \phi^3
	\right.
	\nonumber \\
	&\left.
	-\frac{7}{108\xi_s}\phi^4
	+\frac{\xi_h m_\phi}{\sqrt{2\xi_s}} \phi h^2
	-\frac{\xi_h}{2\xi_s}\phi^2h^2
	\right].
	\label{eq:action_ren}
\end{align}
The Higgs quartic coupling in our system is $\lambda_\mathrm{UV}$, 
and hence it should be smaller than $\sim 4\pi$ for the theory to be perturbative.
Thus, we may require the following condition:\footnote{
	We implicitly assume $\lvert \lambda_h \rvert \lesssim \mathcal{O}(0.1)$.
}
\begin{align}
	\frac{\xi_h^2}{\xi_s} \lesssim 4\pi,
	\label{eq:perturbative}
\end{align}
otherwise it is unreasonable to use the tree-level potential such as Eq.~\eqref{eq:action_ren}.
It means that the theory with non-zero $\xi_s$ is qualitatively different from that with $\xi_s = 0$ even
if we take the limit $\xi_s \rightarrow 0$ because of the presence of the scalaron.
Once we have the scalaron degree of freedom, we need Eq.~\eqref{eq:perturbative}
to keep the theory perturbative at around the vacuum,
while there is no such a constraint if $\xi_s = 0$ from the beginning.
It might be interesting to see that Eq.~\eqref{eq:perturbative} requires the inflaton mass as
\begin{align}
	m_\phi^2 \lesssim \frac{M_P^2}{\xi_h^2} \sim \Lambda_{\mathrm{cut};h}^2,
\end{align}
where the right-hand-side is the cut-off scale of the Higgs inflation model at around the vacuum.
Since we consider the case $\lvert\xi_h\rvert \gg 1$, the inflaton-Higgs quartic coupling is always
in the perturbative region once Eq.~\eqref{eq:perturbative} is satisfied.

We should note that the perturbativity of our system is important 
to safely obtain the SM at the energy scale below $m_\phi$.
In order to see this point, we now derive the infrared (IR) theory from our system
(or the UV theory) below the energy scale of $m_\phi$.
We may define the IR theory as
\begin{align}
	S_\mathrm{IR} =
	\int d^4x \sqrt{-g}
	&\left[ \frac{M_P^2}{2}R - \frac{1}{2}\left(\partial h\right)^2
	-\frac{\lambda_\mathrm{IR}}{4}h^4
	\right].
	\label{eq:action_renIR}
\end{align}
Then, $\lambda_\mathrm{IR}$ may be determined by matching, \textit{e.g.} the scattering process
$hh\rightarrow hh$ in the IR and UV theories.\footnote{
	We consider only the degrees of freedom of $h$ and $\phi$ for simplicity.
} If the UV and IR theories are perturbative, 
we may obtain $\lambda_\mathrm{IR} = \lambda_h$
by comparing the tree level processes in the IR and UV theories.
For more details on this point, see App.~\ref{app:match}.
Thus, the IR theory is nothing but the SM
with the Higgs quartic coupling given by $\lambda_h$.
If the UV theory is strongly coupled, however, 
the tree-level matching does not make sense.
We need to sum an infinite numbers of diagrams to calculate the scattering,
which is a complicated task.
Moreover, it is even non-trivial whether the Higgs remains as an asymptotic state,
and the IR theory might be totally different from the one described 
by Eq.~\eqref{eq:action_renIR}. 
Therefore it is at least secure to consider only the parameter region $\xi_h^2/\xi_s \lesssim 4\pi$.


\section{Metastable electroweak vacuum}
\label{sec:metastable}
In this section, we discuss the two-field dynamics of our system in the case~(c): 
$\xi_h<0,~\xi_s>0$ and $\lambda_h<0$.
It corresponds to the case of the metastable EW vacuum.
Note that 
the current measurement of the top and Higgs masses actually indicates that the Higgs quartic
coupling becomes negative at a high energy region, 
resulting in the metastable EW vacuum~\cite{Sher:1988mj,Arnold:1989cb,Anderson:1990aa,Arnold:1991cv,
Espinosa:1995se,Isidori:2001bm,Espinosa:2007qp,Ellis:2009tp,Bezrukov:2009db,EliasMiro:2011aa,Bezrukov:2012sa,
Degrassi:2012ry,Buttazzo:2013uya,Bednyakov:2015sca}.

We now study the dynamics of the Higgs during and after inflation.
During inflation, the potential minimum for $\tau$ is
\begin{align}
	\tau = \infty,
\end{align}
and hence it corresponds to the usual $R^2$-inflation.
In this case, as long as $\lvert \xi_h \rvert \gg 1$, 
the Higgs stays at $h = 0$ during inflation
since $\tau$ is heavy enough (see Eq.~\eqref{eq:second_deriv1}).
It is nothing but the stabilization mechanism discussed, \textit{e.g.} in 
Refs.~\cite{Espinosa:2007qp,Lebedev:2012sy,Kamada:2014ufa,Espinosa:2015qea,Calmet:2017hja}.
It is well-known that the EW vacuum metastability has some tension 
with high-scale inflation models including the $R^2$-inflation if there
is no coupling between the inflaton and the Higgs 
sectors~\cite{Espinosa:2007qp,Kobakhidze:2013tn,Fairbairn:2014zia,Hook:2014uia,Herranen:2014cua,
Kearney:2015vba,Espinosa:2015qea,East:2016anr}.
However, once we introduce couplings between the inflaton and/or the Ricci scalar and the Higgs,
the EW vacuum is stabilized during inflation since they induce an effective mass of the Higgs.
Here we have explicitly shown that such a stabilization mechanism does work
for the $R^2$-inflation model.

After inflation, however, resonant Higgs production
typically occurs due to the inflaton oscillation, 
and hence the EW vacuum may be destabilized
during the preheating epoch~\cite{Herranen:2015ima,Ema:2016kpf,Kohri:2016wof,Enqvist:2016mqj,Ema:2017loe}.
Here we consider the dynamics of the Higgs after inflation in our system.
Soon after inflation ends, the $\phi^3$ and $\phi^4$ terms as well as the Planck suppressed terms
become negligible due to the cosmic expansion.
In the same way, the Higgs-inflaton quartic coupling becomes less important
than the Higgs-inflaton trilinear coupling.
Therefore we may approximate the action as
\begin{align}
	S 
	\simeq
	\int d^4x \sqrt{-g}
	&\left[ \frac{M_P^2}{2}R - \frac{1}{2}\left(\partial \phi\right)^2 
	- \frac{1}{2}\left(\partial h\right)^2
	\right. 
	\nonumber \\ 
	&\left.
	-\frac{\lambda_\mathrm{UV}}{4}h^4
	-\frac{m_\phi^2}{2}\phi^2
	+\frac{\xi_h m_\phi}{\sqrt{2\xi_s}} \phi h^2
	\right].
	\label{eq:action_meta}
\end{align}
There are two cases depending on the sign of $\lambda_\mathrm{UV}$.
If $\lambda_\mathrm{UV}$ is negative, or $\xi_h^2 < \lvert \lambda_h \rvert \xi_s$,
this system is studied in Ref.~\cite{Enqvist:2016mqj,Ema:2017loe}.\footnote{
	Actually in that study they also include the positive Higgs-inflaton quartic coupling,
	but the trilinear coupling eventually becomes more important than the quartic coupling,
	so we can safely apply their result to our system.
} In this case, the so-called tachyonic preheating occurs since the effective mass squared 
of the Higgs oscillates between positive and negative values~\cite{Dufaux:2006ee}.
As a result, the EW vacuum is destabilized
during the preheating epoch if the trilinear coupling satisfies~\cite{Enqvist:2016mqj,Ema:2017loe}
\begin{align}
	\left\lvert \frac{\xi_h m_\phi}{\sqrt{\xi_s}} \right\rvert \gtrsim \mathcal{O}(10)\times\frac{m_\phi^2}{M_P}.
\end{align}
In order to prevent such a catastrophe, $\xi_h$ must satisfy
\begin{align}
	\left\lvert \xi_h \right\rvert \lesssim \mathcal{O}(10).
\end{align}
Note that the Higgs-inflaton quartic coupling makes things even worse in our case
since it contributes negatively to the effective mass squared of the Higgs for $\xi_h < 0$.

If $\lambda_\mathrm{UV}$ is positive, or $\xi_h^2 > \lvert \lambda_h \rvert \xi_s$,
the situation is more complicated.
The direction $\phi = 0$ considered 
in Refs.~\cite{Herranen:2015ima,Ema:2016kpf,Kohri:2016wof,Enqvist:2016mqj,Ema:2017loe}
is absolutely stable in this case. 
Nevertheless, the potential has an unstable direction $\phi = \xi_h h^2/M_P$
(or more precisely $e^\chi - 1 = \xi_h h^2/M_P$)
since $\lambda_h$ is negative. Fluctuations in this direction might be enhanced
since the inflaton inevitably couples to this direction.
Hence it might be possible that the EW vacuum is destabilized even in this case for large enough $\lvert \xi_h\rvert$,
although we leave a detailed study as a future work.

\section{Summary and discussions}
\label{sec:sum}

In this paper, we have considered the inflationary dynamics of a system with the non-minimal coupling
between the Higgs and the Ricci scalar $\xi_h h^2 R$ as well as the Ricci scalar squared term $\xi_sR^2$.
In such a system, there are two scalar degrees of freedom, 
\textit{i.e.} the Higgs and the scalar part of the metric, or the scalaron.
We have shown that inflation successfully occurs 
in the following three cases: (a)~$\xi_h > 0,~ \xi_s > 0$ and $\lambda_h > 0$,
(b)~$\xi_h < 0,~ \xi_s>0$ and $\lambda_h>0$, and (c)~$\xi_h<0,~\xi_s>0$ and $\lambda_h<0$,
where $\lambda_h$ is the Higgs quartic coupling in the Jordan frame.
We have seen that in every case the inflationary dynamics effectively reduces to a single field one
since the direction orthogonal to the inflaton is heavy enough for $\lvert \xi_h \rvert \gg 1$.
In particular, in the case~(a), the inflaton is a mixture of the Higgs and the scalaron, 
which we call a Higgs scalaron mixed inflation. 
The inflaton potential in this case is given by
\begin{align}
	U	
	= \frac{M_P^4}{4}\frac{1}{\xi_h^2/\lambda_h + \xi_s}\left[1-\exp\left(-\sqrt{\frac{2}{3}}\frac{\phi}{M_P}\right)\right]^2,
	\label{eq:pot_inf_sum}
\end{align}
where $\phi$ is the inflaton and $M_P$ is the reduced Planck mass,
and hence it is consistent well with the CMB observation as long as 
$\xi_h^2/\lambda_h + \xi_s \simeq 2\times 10^{9}$.

We have also addressed the unitarity structure of our system at around the vacuum,
and found that the cut-off scale is as large as the Planck scale.
This is in contrast to the Higgs-inflation where the cut-off scale at around the vacuum
is $M_P/\xi_h \ll M_P$.
Rather, it is similar to the $R^2$-inflation.
Still, the parameters must satisfy $\xi_h^2/\xi_s \lesssim 4\pi$ if we require the perturbativity of our system,
and hence the inflaton mass should be less than $\sim M_P/\xi_h$.

Finally we have briefly discussed the implications of the metastable EW vacuum
to the $R^2$-inflation.
We have explicitly shown that if $\lvert \xi_h \rvert \gg 1$, 
the EW vacuum is not destabilized during inflation albeit it is metastable.
Still, however, it is possible that the EW vacuum is destabilized during the inflaton oscillation
epoch due to a resonant enhancement of the Higgs quanta.
In order to avoid such a catastrophe, we might require $\lvert \xi_h \rvert \lesssim \mathcal{O}(10)$,
although the situation is unclear 
when $\lvert\xi_h\rvert$ is large enough so that $\xi_h^2 \gtrsim \lvert\lambda_h\rvert\xi_s$.
It may be interesting to study further on this respect.

We have several remarks.
First of all, although we have mainly concentrated on the inflationary dynamics in this paper,
the (p)reheating dynamics after inflation is also important.
The inflationary predictions of our system depend
on the reheating temperature thorough the number of e-foldings.
Actually, it is the dynamics after inflation that makes 
the differences of the inflationary predictions 
between the Higgs- and $R^2$-inflation~\cite{Bezrukov:2011gp}.
We leave a detailed study of the reheating dynamics as a future work.

Related to the reheating dynamics, we point out the difference of our system
to the Higgs inflation during the inflaton oscillation epoch.
In general, if there is a large non-minimal coupling between 
the inflaton and the Ricci scalar (without a scalaron) as in the Higgs-inflation,
the inflaton dynamics shows a peculiar behavior called a ``spike''-like feature
during the inflaton oscillation 
epoch~\cite{Ema:2016dny,DeCross:2015uza,DeCross:2016fdz,DeCross:2016cbs},
which was not taken into account in the previous studies~\cite{Bezrukov:2008ut,GarciaBellido:2008ab}.\footnote{
	Actually there is a comment on the energy scale $\sim \sqrt{\lambda_h} M_P$
	related to this spike-like feature in Sec.~3.4 in Ref.~\cite{Bezrukov:2008ut}.
	Nevertheless, possible effects on the preheating dynamics are underestimated there.
} In the Jordan frame, this is because the kinetic term of the inflaton 
(or the Higgs) suddenly changes when the inflaton passes 
the points $\lvert\phi\rvert \sim M_P/\xi_h$,
which has some influence even in the Einstein frame.
In particular, if the inflaton is gauge-charged such as the Higgs,
longitudinal gauge bosons with extremely high-momentum $\sim \sqrt{\lambda_h} M_P$ 
are efficiently produced at the first oscillation so that the unitarity may be violated~\cite{Ema:2016dny}.
Here	it is essential to note that the longitudinal gauge boson mass is different from
the transverse gauge boson one if the symmetry breaking field (the Higgs in our case)
is time-dependent~\cite{Graham:2015rva,Lozanov:2016pac,Ema:2016dny}.
In our case with the scalaron, however, 
such a violent phenomena does not occur thanks to the presence of the scalaron.
Thus, we can trust our system from inflation until the present universe.

It is also valuable to note that some Higgs condensation is unavoidably  produced
in the Higgs scalaron mixed inflation case even if the Higgs is sub-dominant. 
The amplitude of the Higgs condensation at the beginning of the inflaton oscillation
is estimated for the Higgs sub-dominant case as follows.
During inflation, the ratio $s/h^2$ is fixed to be $\lambda_h/\xi_h$, and hence
\begin{align}
	\exp\left(\sqrt{\frac{2}{3}}\frac{\phi}{M_P}\right) - 1 = \frac{\xi_h h^2 + \xi_s s}{M_P^2}
	\simeq \frac{\lambda_h \xi_s}{\xi_h}\frac{h^2}{M_P^2}.
\end{align}
The left-hand-side is of order unity at the beginning of the inflaton oscillation,
and thus the amplitude of the Higgs condensation at that time is estimated as
\begin{align}
	h_\mathrm{osc} \sim \sqrt{\frac{\xi_h}{\lambda_h \xi_s}}M_P \sim 10^{-5}\sqrt{\frac{\xi_h}{\lambda_h}}M_P,
\end{align}
where we substitute $\xi_s \simeq 2\times 10^{9}$ from the CMB observation in the last similarity.
Although the dynamics of the Higgs condensation after inflation is non-trivial
due to the couplings to the inflaton,
it might have some phenomenological consequences such as 
the spontaneous leptogenesis~\cite{Kusenko:2014lra} and
the gravitational wave~\cite{Figueroa:2014aya,Figueroa:2016ojl}.
Note that in our case the Higgs has no isocurvature perturbation 
as opposed to the case considered 
in Refs~\cite{DeSimone:2012qr,Kunimitsu:2012xx,Choi:2012cp,Enqvist:2013kaa}
since $\tau$ is massive during inflation.
\\
\vspace{0.3cm}

\noindent
\textbf{Note Added:} While finalizing this paper, another paper~\cite{Wang:2017fuy} 
was submitted to the arXiv that calculated the curvature perturbation in the same system.

\section*{Acknowledgments}
{\small
YE thanks Kazunori Nakayama for useful discussions.
YE also acknowledges Oleg Lebedev and the members of the theoretical high energy physics group 
in the University of Helsinki for their hospitality, where some part of this work was done.
This work was supported by the JSPS Research Fellowships for Young Scientists
and the Program for Leading Graduate Schools, MEXT, Japan.
}
\appendix

\section{Redundancy of dual description}
\label{app:eqv}
In this appendix, we comment on redundancy of dual description of Eq.~\eqref{eq:action_J1}. 
Instead of Eq.~\eqref{eq:action_J2}, we may consider the following action:
\begin{align}
	S = \int d^4x \sqrt{-g_J}
	&\left[
	\frac{M_P^2}{2}\left(1+\frac{\xi_h' h^2 + \xi_s' s}{M_P^2}\right)R_J \right. \nonumber \\
	&\left.
	- \frac{\xi_s''}{4}s^2-\frac{\lambda_{sh}}{2}s h^2
	-\frac{1}{2}g^{\mu\nu}_J\partial_\mu h\partial_\nu h - \frac{\lambda_h'}{4}h^4
	\right],
	\label{eq:action_Japp}
\end{align}
where the parameters $\xi_h'$, $\xi_s'$, $\xi_s''$ and $\lambda_h'$ are in general different
from $\xi_h$, $\xi_s$ and $\lambda_h$.
As long as they satisfy
\begin{align}
	\xi_h' - \frac{\xi_s'}{\xi_s''}\lambda_{sh} &= \xi_h,~~~
	\frac{\xi_s'^2}{\xi_s''} = \xi_s,~~~
	\lambda_h' - \frac{\lambda_{sh}^2}{\xi_s''} = \lambda_h,
	\label{eq:param_eqv}
\end{align}
the action~\eqref{eq:action_Japp} reproduces Eq.~\eqref{eq:action_J1}
after integrating out the scalaron $s$, and hence they are actually redundant.
This redundancy corresponds to the shift and the rescaling of the scalaron $s$,
and the physics does not depend on this ambiguity of the dual description.
We can also see this in the Einstein frame.
By following the same procedure we obtained Eq.~\eqref{eq:action_E},
we can show that the Einstein frame action obtained from Eq.~\eqref{eq:action_Japp}
is exactly Eq.~\eqref{eq:action_E} thanks to Eq.~\eqref{eq:param_eqv}.
In the main text, we have chosen the parameters as
\begin{align}
	\xi_h' = \xi_h,~~~
	\xi_s' = \xi_s'' = \xi_s,~~~
	\lambda_{sh} = 0,~~~
	\lambda_h' = \lambda_h,
\end{align}
which of course satisfy Eq.~\eqref{eq:param_eqv}.

\section{Matching}
\label{app:match}

\begin{figure}[t]
\begin{center}
\includegraphics[scale=0.33]{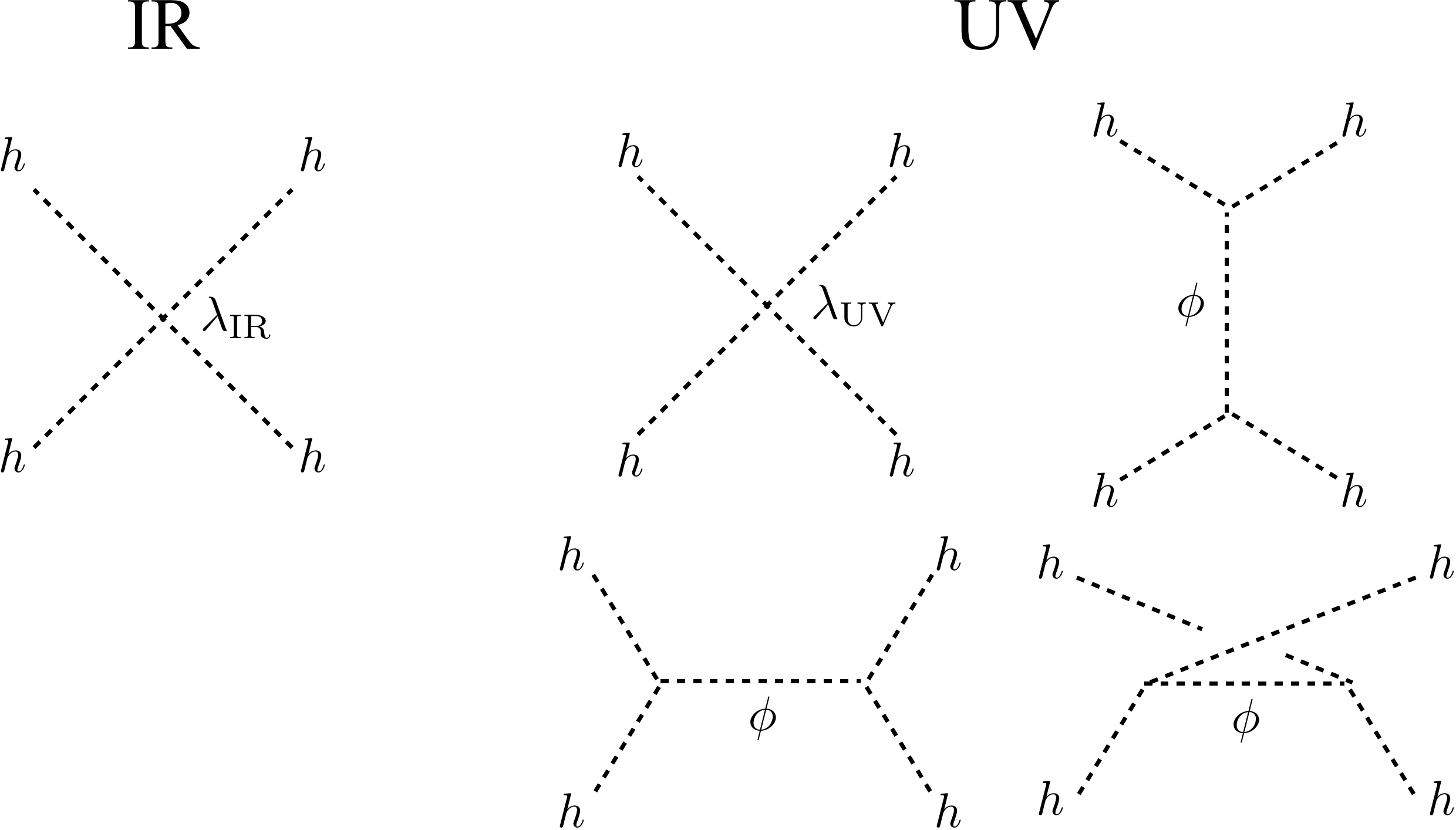}
\end{center}
\caption { The Feynman diagrams that contribute to the scattering process $hh\rightarrow hh$
		at the tree level in the IR theory (left) and the UV theory (right), respectively.
}
\label{fig:match}
\end{figure}

In this appendix, we explain how to obtain the relation $\lambda_\mathrm{IR} = \lambda_h$.
In order to express the coupling in the IR theory in terms of those in the UV theory,
we may compare the tree-level scattering amplitude of the process $hh\rightarrow hh$.
The corresponding diagrams in the IR and UV theories are shown in Fig.~\ref{fig:match}.
In the IR theory, the tree-level amplitude is given by
\begin{align}
	i\mathcal{M}_\mathrm{IR} = -6i\lambda_\mathrm{IR},
\end{align}
where the numerical factor comes from the permutation of the external Higgs particles.
In the UV theory, it is given by
\begin{align}
	i\mathcal{M}_\mathrm{UV} 
	&= -6i\lambda_\mathrm{UV}
	-\frac{2\xi_h^2 m_\phi^2}{\xi_s}
	\left(\frac{i}{s-m_\phi^2}+\frac{i}{t-m_\phi^2}+\frac{i}{u-m_\phi^2}\right) \nonumber \\
	&\simeq -6i\left(\lambda_\mathrm{UV} - \frac{\xi_h^2}{\xi_s}\right) 
	+ \mathcal{O}\left(\frac{k^2}{m_\phi^2}\right),
\end{align}
where the second term comes from the inflaton-Higgs trilinear coupling with 
$s, t$ and $u$ being the Mandelstam variables.
In the last line, we have taken only
leading order terms in $\mathcal{O}(k^2/m_\phi^2)$
with $k$ being the typical momentum of the scattering.
Thus, by matching the scattering amplitude, we obtain
\begin{align}
	\lambda_\mathrm{IR} = \lambda_\mathrm{UV} - \frac{\xi_h^2}{\xi_s} = \lambda_h.
\end{align}
Note that this procedure is reasonable only when the IR and UV theories are perturbative.

\bibliographystyle{apsrev4-1}
\bibliography{ref}

\end{document}